# Quantum size effects in ultra-thin YBa$_2$Cu$_3$O$_{7-x}$ films

M. Lyatti[1,2*], I. Gundareva[1,2], T. Röper[1,2], Z. Popovic[3], A.R. Jalil[1,2], D. Grützmacher[1], T. Schäpers[1,2]

[1] Peter Grünberg Institut (PGI-9), Forschungszentrum Jülich, 52425 Jülich, Germany

[2] JARA-Fundamentals of Future Information Technology, Jülich-Aachen Research Alliance, Forschungszentrum Jülich and RWTH Aachen University, Germany

[3] University of Belgrade, Faculty of Physics, Studentski trg. 12, 11001 Belgrade, Serbia

**Abstract**

The d-wave symmetry of the order parameter with zero energy gap in nodal directions stands in the way of using high-temperature superconductors for quantum applications. We investigate the symmetry of the order parameter in ultra-thin YBa$_2$Cu$_3$O$_{7-x}$ (YBCO) films by measuring the electrical transport properties of nanowires and nanoconstrictions aligned at different angles relative to the main crystallographic axes. The anisotropy of the nanowire critical current in the nodal and antinodal directions reduces with the decrease in the film thickness. The Andreev reflection spectroscopy shows the presence of a thickness-dependent energy gap that does not exist in bulk YBCO. We find that the thickness-dependent energy gap appears due to the quantum size effects in ultra-thin YBCO films that open the superconducting energy gap along the entire Fermi surface. The fully gapped state of the ultra-thin YBCO films makes them a promising platform for quantum applications, including quantum computing and quantum communications.

**Main**

Conventional low-temperature superconductors are widely used for quantum applications because of an s-wave pairing symmetry where the energy gap exists along the entire Fermi surface providing an exponentially low number of unpaired quasiparticles at low temperatures. High-temperature (high-$T_c$) cuprate superconductors have significantly higher critical temperatures and larger superconducting energy gaps which is beneficial for many applications. However, optimally-doped bulk cuprate superconductors have d$_{x2-y2}$-wave pairing symmetry where the energy gap vanishes in nodal directions[1, 2], as shown in Figure 1a. The gapless state reduces the significance of high-$T_c$ superconductors for quantum applications because the unpaired quasiparticles in these superconductors exist even at zero temperature.

Nonetheless, there are a number of experimental results including the parity effect in YBa$_2$Cu$_3$O$_{7-x}$ (YBCO) nanoparticles[3], the signatures of s-wave pairing in ultra-thin Bi$_2$Sr$_2$CaCu$_2$O$_{8+x}$ Josephson junctions[4] and disordered YBCO films[5], the single-photon response of the high-$T_c$ nanowires[6, 7], and the high quality factor of YBCO nanowires[8] that cannot be explained by the d-wave pairing symmetry. The nodeless energy gap has been reported for optimally doped YBCO, underdoped $La_{2-x}Sr_xCuO_4$, and $Bi_2Sr_{2-x}La_xCuO_{6+\delta}$ films[9-11]. Theoretically, it was proposed that the scattering[12], disorder[13], or large superconducting phase fluctuations[14, 15] may result in the fully gapped state of the d-wave superconductor. All this evidence gives hope that it is feasible to build fully gapped devices from the high-$T_c$ superconductors which can be used for quantum applications. However, there is still no recipe for preparing the fully gapped state in the optimally- or overdoped d-wave superconductor.

* m.lyatti@fz-juelich.de   1

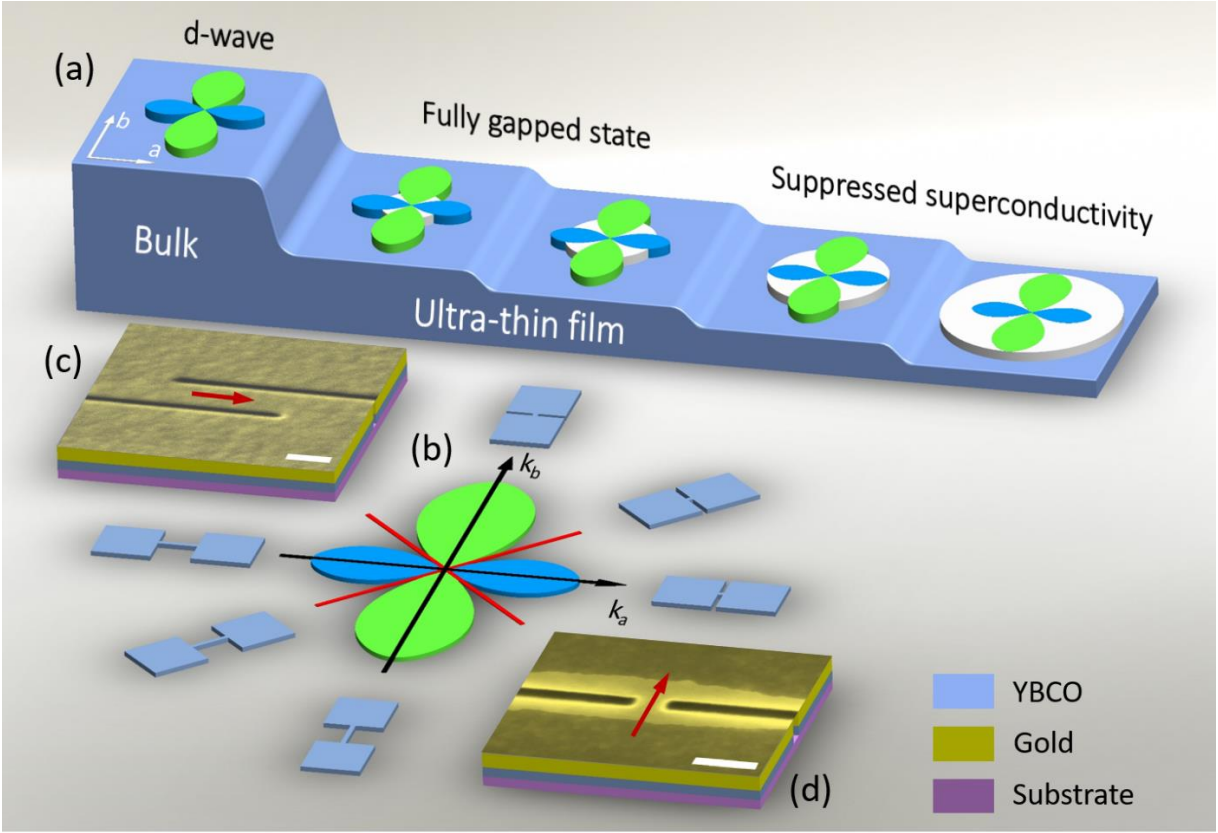

**Figure 1| Energy gap in the nanoscale YBCO film. a.** Evolution of the energy gap in YBCO with decreasing film thickness. The superconducting energy gap in bulk YBCO is colored in blue and green. The confinement energy gap is colored in white. **b**. Orientation of the nanowires and nanoconstriction with respect to the YBCO crystallographic axes. Red lines show nodal directions. **c, d**. SEM micrographs of YBCO nanowire and nanoconstriction. Red arrows show the direction of the current flow. Scale bars correspond to 200 nm.

Following the observation of the fully gaped state in the ultra-thin YBCO nanowires[8], we fabricate nanowires and nanoconstrictions and use them to investigate the order parameter symmetry in ultra-thin YBCO films and elucidate the nature of the order parameter symmetry transformation. Investigations of the nanoconstrictions and nanowires properties complement each other. Relative values of the energy gap with a high angular resolution can be obtained from the measurements of the nanowire transport characteristics. While the Andreev spectra of nanoconstriction provide information on the absolute magnitude of the energy gap in a wide range of angles.

**Study of order parameter symmetry with nanowires**

We fabricate 530-nm-long nanowires with an effective width of 70-100 nm from the over-doped ultra-thin YBCO films with a thickness $d$ in the range of 7 – 11.6 nm (6-10 unit cells (u.c.)) deposited on a (100) SrTiO$_3$ substrate. The c-axis-oriented YBCO films have a roughness of ±1 u.c.[16]. On each substrate, we pattern 13-27 nanowires along the substrate diagonal and edges, which correspond to the nodal and antinodal directions of YBCO, respectively, as schematically shown in Figure 1b. The a- and b-axis directions of the YBCO film can be identified from the XRD measurements (see Supplementary information). The representative SEM micrograph of the nanowire shaped by two cuts using focused ion beam (FIB) is demonstrated in Figure 1c. The nanowires are covered with a 15-nm-thick gold layer deposited *in situ* to get the same boundary conditions as in nanoconstrictions, presented below, where the gold layer prevents overheating at high voltage biases.

We measure the current-voltage (*IV*) characteristic of the current-biased nanowire at a temperature $T$ = 4.2 K and determine the critical current $I_c$ using a 10 µV voltage threshold. Most *IV* curves demonstrate a voltage switching at a current above the critical current with the small current hysteresis due to the phase slippage by the kinematic vortex motion [17-20] (see the Supplementary information). An average critical current density of the nanowires oriented along one substrate edge is higher than that of the nanowires oriented along the perpendicular substrate edge as previously observed for the twin-free or partially twinned YBCO films [21]. Therefore, we assign the nanowires with the lower and higher critical current densities to the **a**- and **b**-axis directions, respectively.

The critical current density is calculated as $J_c=I_c/W_{eff}d_{eff}$ using the effective width $W_{eff}$ = $W$ – 140 nm instead of geometric width $W$, as outlined in our previous work [17], and the effective thickness $d_{eff}$ = $d$ -2 u.c. taking into account the non-superconducting YBCO layers at the YBCO film interfaces [16]. We estimate a reduction in critical current density due to current crowding at the ends of the nanowires of less than 0.5% because of the large turnaround radius [22].

The thickness dependences of the average critical current density $<J_{c0}>$ in the **a**-axis direction (0°), $<J_{c90}>$ in the **b**-axis direction (90°), and $<J_{c45}>$ in the nodal direction (45°) are plotted in Figure 2. The average critical current density $<J_c>$ in the antinodal directions significantly reduces with the decrease of the film thickness preserving the same $<J_{c0}>/<J_{c90}>$ ratio of 0.7 for the 7-10-u.c.-thick nanowires as shown in the inset in Figure 2. The measured $<J_{c0}>/<J_{c90}>$ ratio is close to the anisotropy of the London penetration depth $\lambda_b/\lambda_a$ and superconducting energy gap in the single-crystal YBCO samples [23, 24].

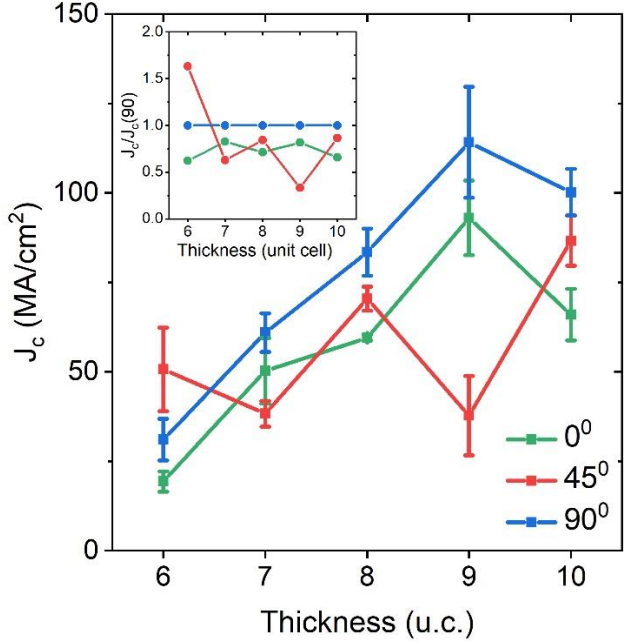

**Figure 2|** **The thickness dependence of the critical current density $J_c$ of the nanowires oriented in different directions with respect to the a-axis at $T$ = 4.2 K. Error bars represent one standard deviation.** Inset shows the dependence of the normalized critical current density on the nanowire orientation relative to the a-axis.

The critical current density in the nodal direction is comparable to those in the antinodal ones which differs from the bulk samples, where $J_{c45}/J_{c0}$ and $J_{c45}/J_{c90}$ ratios are about a few percent [25, 26]. Remarkably, the $<J_{c45}>$ thickness dependence (inset in Figure 2) is different from those in the antinodal directions. The critical current density in the nodal direction does not demonstrate a monotonic decrease but rather scatters between 38 and 86 MA/cm² which might be a signature of size-dependent oscillations of the density of states at the Fermi energy level [27]. The large values of the nodal critical current density are clear evidence of the non-zero nodal energy gap in the ultra-thin YBCO films, while the nonmonotonic $<J_{c45}>$ thickness dependence indicates that the origin of this gap differs from that of the intrinsic energy gap in YBCO.

**Study of order parameter symmetry with nanoconstrictions**

After getting evidence of the non-zero energy gap in the nodal directions, we now turn to the question of the absolute value of this gap. To answer this question, we fabricate nanoconstrictions from 5-9-u.c.-thick YBCO films and measure their electrical transport characteristics in the 4.2 – 90 K temperature range. On each substrate, we pattern 7 or 13 nanoconstrictions in the nodal and antinodal directions, as schematically shown in Figure 1b.

At currents above the critical current, a superconducting constriction can be modeled as a superconductor-normal metal-superconductor (SNS) junction where the voltage is developed across the dissipative neck region [28, 29]. Within the framework of the simplified theoretical approach to SNS junctions, each quasiparticle undergoes multiple Andreev reflections (MAR) before it is scatteres or leaves the pair potential well. If a quasiparticle undergoes $n$ Andreev reflections (AR), then $ne$ charges are transferred through the NS boundary in addition to the initial one, and the SNS junction current is enhanced due to AR. Here $n$ is an integer and $e$ is an electron charge. As consequences of MAR, the *IV* curves and especially the conductance curves show many nonlinear structures [30-32].

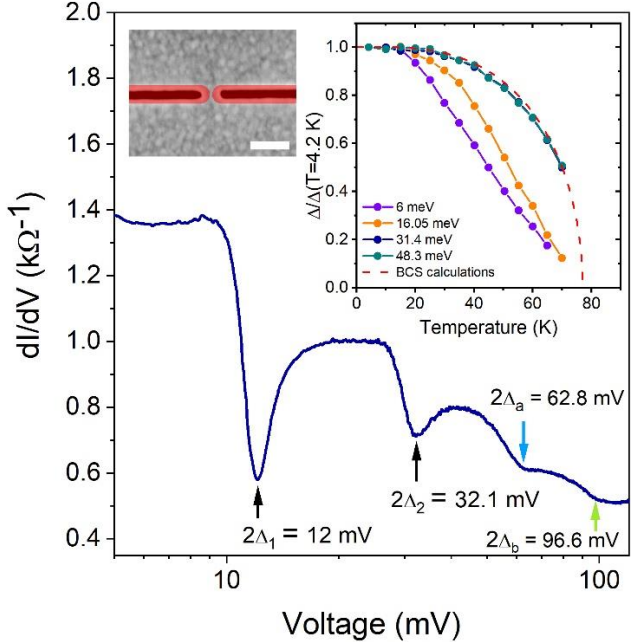

**Figure 3| Differential conductance of the 6-u.c.-thick ultra-narrow YBCO nanoconstriction at 4.2 K.** Blue, green, and black arrows indicate the position of the conductance steps. Inserts: left. SEM image of the ultra-narrow nanoconstriction just after the FIB milling. The non-superconducting YBCO is colored in red. The scale bar corresponds to 200 nm. Right. the temperature dependences of the normalized energy gaps. The dashed line shows the BCS theory prediction.

At higher voltages $V \gtrsim \Delta/2e$, the number of AR becomes energy limited, decreasing with increasing energy that the quasiparticle received per AR and only a few isolated dips occur in the conductance curve[32]. In this high-voltage region, each current enhancement due to AR leads to the corresponding step-wise increase of the SNS junction conductance. According to these theoretical predictions, the measurements of the AR spectrum provide information on the magnitude of the superconducting energy gaps in the nanoconstriction electrodes[29-33].

The YBCO has large superconducting energy gaps [23] that make the conductance measurements up to voltages $V \geq 2\Delta/e \approx 90$ mV challenging because of the constriction overheating. First, we use FIB milling (see Methods) to fabricate ultra-narrow constrictions with very high resistances from bare YBCO films capped with the amorphous YBCO layer that do not show overheating even at voltages above 100 mV. The number of such ultra-narrow constrictions with a measurable critical current was limited by several samples because of a low fabrication yield. The differential conductance *dI/dV* and the SEM micrograph of the high-resistance 6-u.c.-thick YBCO nanoconstriction with the $I_c$ = 11 µA at *T*=4.2 K are presented in Figure 3 and the left insert in Figure 3, respectively. The estimated width of this constriction is 3-6 nm which is comparable to the a-b plane coherence length of YBCO $\xi_{ab}$ = 1.3 nm [34]. The *dI/dV* curve has a well-defined ladder-like structure due to AR in the normal-state region of the nanoconstriction [29, 32, 33]. We exclude the appearance of the conductance features due to the multiple phase-slip centers because the nanowires are significantly shorter than the electric field penetration depth in YBCO[17]. The sharpness and shape of the AR features depend on many parameters including scattering, constriction length, quasiparticle mean-free path, and others. Our numerical simulations show that for the given barrier properties, the sharpness of the conductance step is an indicator of the energy gap anisotropy in the electrodes. The sharper the step the lower the anisotropy. The positions of the smeared steps at voltages $V \approx 62.8$ mV and 96.6 mV shown by blue and green arrows are very close to the expected values of $2\Delta/e$ corresponding to the highly-anisotropic $\Delta_a$ and $\Delta_b$ energy gaps in the **a**- and **b**-axis directions in bulk YBCO, respectively [23]. Their temperature dependences overlap and are close to the prediction of the Bardeen–Cooper–Schrieffer (BCS) theory, as shown in the right inset in Figure 3 (the *dI/dV* dependencies measured in the 4.2 – 80 K temperature range are available in the Supplementary information). The simultaneous presence of the conductance steps originating from the energy gaps in both antinodal directions can be explained by the very wide cone of the quasiparticle trajectories inside the constriction.

The conductance steps at 12 mV and 32.1 mV have larger amplitude and different temperature dependence compared to those at $V = 2\Delta_a/e, 2\Delta_b/e$. The temperature dependencies of the current position of these steps are different from that of the critical current, as shown in the Supplementary information, giving confidence that these features are not critical current features. These steps are sharper and, therefore, the corresponding energy gaps with $\Delta_1(4.2K) = 6$ meV and $\Delta_2(4.2K) = 16.05$ meV have smaller anisotropy compared to $\Delta_a$ and $\Delta_b$ energy gaps. We believe that the energy gap $\Delta_1(4.2K) = 6$ meV belongs to the deoxygenated YBCO adjacent to the constriction area because of the characteristic tail at high temperatures. While the energy gap with $\Delta_2(4.2K) = 16.05$ meV appears in the over-doped YBCO. AR spectra of the ultra-narrow constrictions provide clear evidence of new energy gaps in ultra-thin YBCO films that are not inherent to the bulk YBCO. However, the current in the ultra-narrow nanoconstrictions may flow through the oxygen-depleted YBCO regions which complicates the analysis.

To overcome this problem, we measure additionally 100-nm-wide YBCO nanoconstrictions covered *in situ* by a 15-20-nm-thick gold layer. The 100-nm-wide constrictions are wide enough to avoid electrical transport across the degraded YBCO in the constriction area. The gold layer removes the heat from the constriction area and prevents the normal-state region in the constriction from expanding. These Au/YBCO nanoconstrictions are fabricated with electron-beam lithography and ion-beam etching, as described in the Methods. A representative SEM micrograph of the nanoconstriction is shown in Figure 1d.

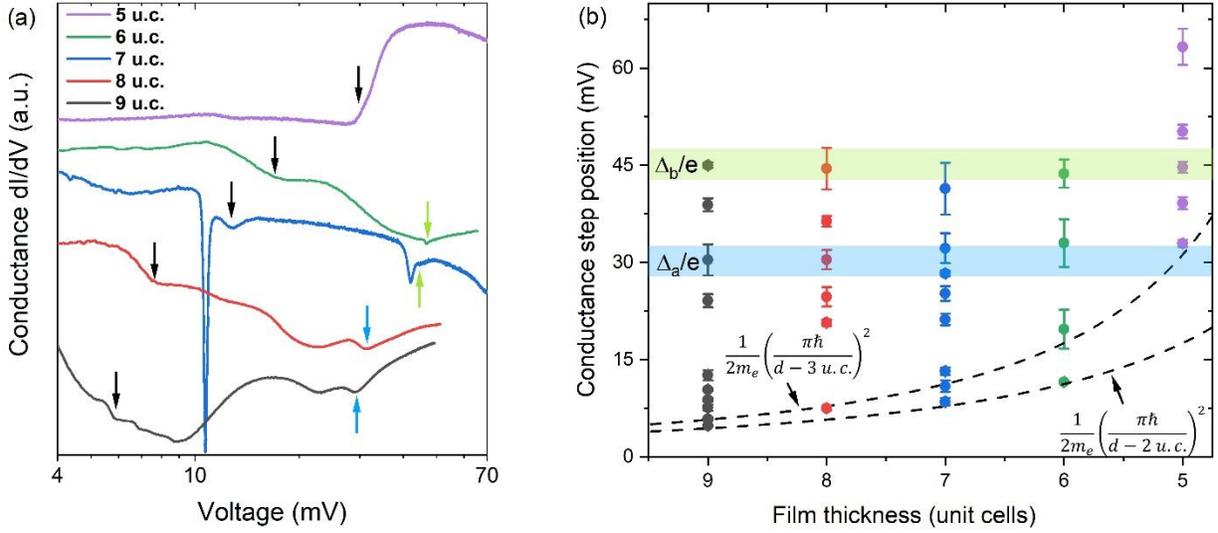

**Figure 4| Energy gaps in ultra-thin YBCO films of various thicknesses.** a. Representative differential conductance of 100-nm-wide Au/YBCO nanoconstrictions of various thicknesses. Curves are shifted along the Y-axis for convenience of their presentation. Blue, green, and black arrows indicate positions of the conductance steps induced by $\Delta_a$, $\Delta_b$, and energy gap not inherent to bulk YBCO, respectively. b. Thickness dependence of the positions of the conductance steps for the Au/YBCO nanoconstrictions. Error bars represent one standard deviation.

Representative differential conductance curves of 100-nm-wide and 5–9-u.c.-thick Au/YBCO nanoconstrictions are shown in Figure 4a. Except for the thinnest nanoconstrictions, the differential conductance demonstrates steps similar to those observed for the ultra-narrow constriction without gold capping layer. We clearly identify the conductance steps at high voltages that correspond to the bulk bulk YBCO energy gaps. These steps, which appear at voltages $V = \Delta_a/e \approx 30$ mV and $V = \Delta_b/e \approx 45$ mV, are indicated in Figure 4a by blue and green arrows, respectively. In addition, we observe conductance steps at lower voltages indicated in Figure 4a by black arrows. Some of the conductance steps are accompanied by very sharp dips as in the case of the 7-u.c.-thick nanoconstriction. These sharp dips can occur in the current bias regime due to the S-type nonlinearity near the $V = 2\Delta/ne$, as shown in the Supplementary information. The conductance of the 5-u.c.-thick nanoconstrictions is almost constant below the voltage of 30 mV and then increases sharply. This behavior is qualitatively different from the Andreev-reflection-like type of conductance observed for thicker nanoconstrictions. The change in the conductance behavior is accompanied by a significant drop in the critical current density by one order of magnitude in one antinodal direction and by two orders of magnitude in the other antinodal and nodal directions.

We plot the positions of the steps found at the differential conductance of the 100-nm-wide Au/YBCO nanoconstrictions of various thicknesses in Figure 4b. The region corresponding to the $\Delta_a/e$ and $\Delta_b/e$ are colored in blue and green, respectively. The conductance steps at voltages around 20 mV can be assigned as $2\Delta_a/3e$ or $\Delta_b/2e$. An accurate MAR analysis is challenging because $\Delta_b \approx 1.5\Delta_a$, therefore, a series of MAR harmonics originating from $\Delta_a$ intersects with a series of MAR harmonics originating from $\Delta_b$. However, the low-voltage conductance steps shown in Figure 4a by black arrows belong neither to $2\Delta_a/ne$ series nor to $2\Delta_b/ne$ series. These conductance steps are caused by the energy gaps which are not inherent to the bulk YBCO. The energies of these low-energy gaps are smaller for the thicker films and higher for the thinner ones. The positions of the conductance steps are limited from below by $(1/d)^2$-type dependence which is shown in Figure 4b by dashed lines.

**Quantum size effects in nanoscale YBCO films**

The $(1/d)^2$-type dependence of the energy gap on the device size is typical for the quantum size effects (QSE) which occur when at least one dimension of the device is comparable with the Fermi wavelength $\lambda_F = h/mv_F$, where $h$ is the Planck constant, $m$ is the charge carrier mass, and $v_F$ is the Fermi velocity. QSE are well known for semiconducting nanodevices because of the large Fermi wavelength [35]. The observation of QSE in metals is more challenging because of their very short Fermi wavelength $\lambda_F \approx 5$ Å. QSE can induce a pseudo gap and change the conductivity type of the atomically-flat nanoscale semimetal [36-38] and metal films [39-41] from metallic to semiconducting. A change in the density of states in the nanoscaled BCS superconductors due to the QSE results in the change in a superconducting energy gap value and, consequently, in the change in the critical temperature [27, 42-48].

The Fermi velocity in high-$T_c$ superconductors is an order of magnitude lower than in metallic superconductors resulting in a larger Fermi wavelength. We estimate the Fermi wavelength in YBCO as $\lambda_F(YBCO) \approx 2$ nm using $m \approx 2m_e$ [49, 50] and $v_F = 2·10^5$ m/sec [51], which is comparable with the total thickness of studied films $d = 5.8 – 10.5$ nm. Therefore, the superconducting properties of our ultra-thin YBCO films which have very low surface roughness have to be affected by QSE.

QSE modify the magnitude of the superconducting energy gap in s-wave superconductors but do not change the order parameter symmetry. In contrast to s-wave superconductors, QSE have a more significant effect on the properties of d-wave superconductors. The bulk d-wave superconductors have nodes where the superconducting energy gap vanishes, as shown in Figure 1a. An opening of the energy gap in the nodal direction, as it is illustrated in a simplified way at the second and third steps from the left side in Figure 1a, brings YBCO into the fully gapped state with an exponentially low number of quasiparticles at low temperatures.

The thickness dependence of the new gap in the ultra-thin YBCO films is in good quantitative agreement with the well-known expression for the confinement energy in the quantum well $E_g = [\hbar\pi/d]^2/2m$ (1) which is shown in Figure 4b by dashed lines. We plot two $E_g$-dependencies which account for the film roughness of 1 u.c. and nonsuperconducting layer at the YBCO-substrate interface. The upper and the lower dashed lines correspond to $E_g = [\hbar\pi/(d\text{-}3 \text{ u.c.})]^2/2m_e$ and $E_g = [\hbar\pi/(d\text{-}2 \text{ u.c.})]^2/2m_e$, respectively. The changes in the superconducting properties become even more drastic when the magnitude of the confinement gap approaches the value of the superconducting energy gap, as shown at the rightmost steps in Figure 1a. The superconducting order parameter is strongly dependent on the number of the single-electron states inside the Debye "window" around the Fermi level [52]. The increasing confinement energy gap first reduces the number of states that can be used for the quasiparticle pairing resulting in the decrease of the critical current density, as it is confirmed by the experimental results for the nanowires in Figure 2. When the magnitude of the confinement gap approaches the value of the Debye energy, which is $\hbar\omega_D \approx 40$ meV $\approx \Delta_b$ in YBCO [53], the critical current density is strongly suppressed and the conductance behavior is dominated by the confinement energy gap as it is observed for the YBCO nanoconstrictions with the total and effective thickness of 5 and 3 u.c., respectively.

The fully gapped state due to QSE explains many unusual experimental results with nanoscale cuprate superconductors which contradict the d-wave symmetry. Finally, we compare our findings with the preceding experimental results. In our previous work, we estimated the superconducting energy gap for 4-u.c.-thick nanowire as 17 meV [8] which is in excellent agreement with the equation (1). Gustafsson et al. [3] reported on the fully gapped state of the 200x200x100 nm$^3$ YBCO nanoparticle with Δ = 18 μeV at zero magnetic field which is again in good quantitative agreement with $E_g$ = 10-40 μeV calculated with equation (1). The increase of the superconducting energy gap of YBCO nanoparticle in strong magnetic field observed by Gustafsson et al. [3] can be explained by the increase of the confinement energy gap in strong magnetic fields [54].

**Conclusions**

We investigate the superconducting properties of ultra-thin YBCO films and find that quantum size effects open the energy gap in the nodal direction and turn the d-wave superconductor into the fully gapped state with the magnitude of the nodal energy gap given by the confinement energy $E_g = (\hbar\pi/d)^2/2m_e$. The nanoscale YBCO film can be considered as a quantum-engineered superconductor where the superconducting gap is controlled by quantum effects. The fully gapped state paves the way for nanoscale d-wave superconductors towards many quantum applications including quantum computing and single-photon detection.

**Methods**

**Nanostructure fabrication.** YBCO nanowires and nanoconstrictions were fabricated from a 5.9- 11.6 nm (5-10 unit cell) thick epitaxial YBCO films deposited on a $TiO_2$-terminated (100) single-crystal $SrTiO_3$ substrate by dc sputtering at high (3.4 mbar) oxygen pressure and substrate heater temperature of 950°C. One unit cell corresponds to the YBCO lattice parameter in **c**-direction which is 11.62 Å. Substrate edges are aligned along (010) and (001) crystollagraphic planes. The film thickness was controlled by the deposition time. The deposition rate of 1.0 nm/min was measured by XRR and atomic-force microscope measurements. After the deposition, the films were annealed in oxygen at a pressure of 800 mbar and substrate heater temperature of 500°C. The YBCO film deposition is described in more detail elsewhere [16]. The structural properties of the ultra-thin YBCO films are described in the Supplementary information. After the annealing, YBCO films were *in situ* capped either with a 6-nm-thick amorphous YBCO layer deposited at room temperature in the case of the ultra-narrow nanoconstrictions or by the 15-20 nm-thick gold layer deposited at $T$ = 100°C by magnetron dc sputtering in argon at a pressure of $5 \cdot 10^{-3}$ mbar. We do not see any effect of the gold layer on the nanowire or nanoconstriction superconducting properties. The 100-nm-thick Au contact pads were deposited *ex situ* using room temperature dc magnetron sputtering with a shadow mask. Following contact pad deposition, the nanostructures were fabricated in a two-step process. In the first step, 5-µm-wide microbridges oriented along [100] and [010] crystallographic axes of the $SrTiO_3$ substrate were patterned using an optical UV contact lithography with a PMMA resist and wet chemical etching in a Br-Ethanol and $I_2$-NaI-Ethanol solutions. Optical micrographs and electrical parameters of the microbridges are available in the Supplementary Information. The electrical characteristics of the selected microbridges on each substrate were measured before nanopatterning to ensure the film quality. In the second step, thirteen 530-nm-long Au/YBCO nanowires or ultra-narrow high-resistance YBCO nanoconstrictions capped with the amorphous YBCO layer were fabricated across the microbridges with two cuts made with FIB milling using an Au/PMMA protective layer. In order to increase the number of nanowires that can be placed at the same substrate and measured with the existing contact system, we also pattern additional fourteen nanowires across the current leads. The Au/PMMa protective layer was removed in acetone before electrical measurements. The width of each nanowire was measured using SEM. More details on the FIB patterning process can be found in Ref. 18. The 100-nm wide and the 40-50-nm-long Au/YBCO nanoconstrictions were made using inverse process. In the first step, the nanostructures were fabricated using the ion-beam etching in argon through the 80-nm-thick CSAR62 resist mask patterned by the electron-beam lithography. In the second step, the microbridges were defined by an optical UV contact lithography and wet chemical etching using the alignment markers fabricated during the first step. In total 158 nanowires and more than 75 nanoconstrictions have been fabricated.

**Experimental setup.** The experimental setup was based on a liquid helium storage Dewar insert filled with He exchange gas. The temperature above 4.2 K was maintained with the resistive heater controlled by a Lake Shore temperature controller 335. We used battery-operated low-noise analog electronics to sweep the bias current and amplify the voltage across the nanostructure. The differential resistance of the nanostructure was measured with lock-in amplifier 7265 (Signal Recovery) at 10 kHz modulation frequency.

**Data availability**

The data that support the findings of this study are available from the corresponding authors upon reasonable request.

**Acknowledgements**

Authors are grateful to A. Golubev for stimulating discussions. The FIB nanopatterning was performed in the Ernst Ruska-Centre of Forschungszentrum Jülich within the project FZJ-PGI-9-LM1. The work of Z.P. was supported by Serbian Ministry of Science, Technological Development and Innovation, Project No. 451-03-47/2023-01/200162. I.G. was supported by the Deutsche Forschungsgemeinschaft (DFG, German Research Foundation) under Germany's Excellence strategy – Cluster of Excellence Matter and Light for Quantum computing (ML4Q) EXC 2004/1 390534769.

**Author contributions**

M.L. and T.R. fabricated the nanostructures and performed the measurements, I.G. fabricated the films and performed the optical lithography, Z.P. performed the numerical simulations of the Andreev reflection spectra, and A.R.J made XRD measurement and analysed data. All authors co-wrote the paper.

**Competing interests**

The authors declare no competing interests.

## Supplementary information

## Quantum size effects in ultra-thin YBa$_2$Cu$_3$O$_{7-x}$ films

M. Lyatti[1,2*], I. Gundareva[1,2], T. Röper[1,2], Z. Popovic[3], A.R. Jalil[1,2], D. Grützmacher[1], T. Schäpers[1,2]

[1] Peter Grünberg Institut (PGI-9), Forschungszentrum Jülich, 52425 Jülich, Germany
[2] JARA-Fundamentals of Future Information Technology, Jülich-Aachen Research Alliance, Forschungszentrum Jülich and RWTH Aachen University, Germany
[3] University of Belgrade, Faculty of Physics, Studentski trg. 12, 11001 Belgrade, Serbia

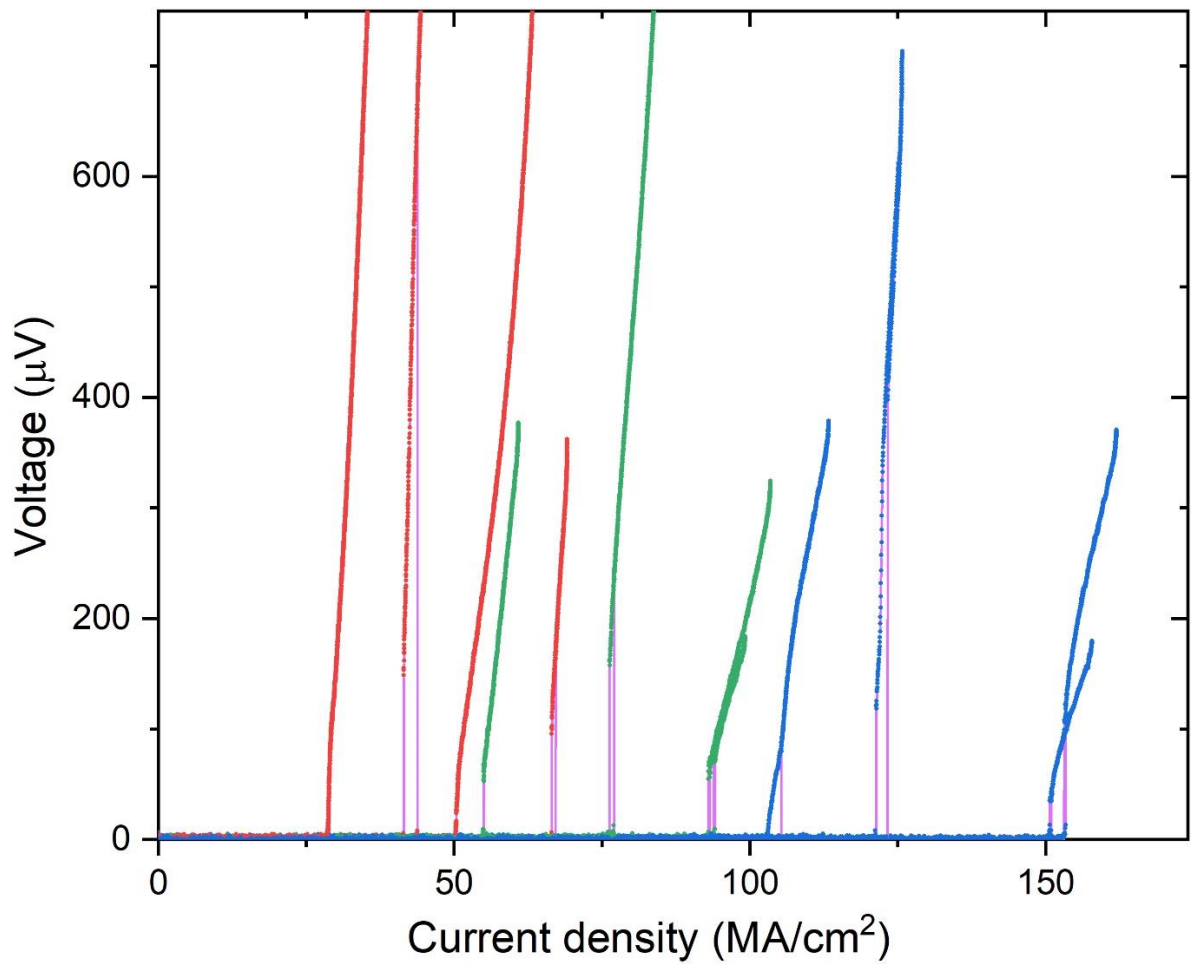

**Figure S1|** *IV* curves of twelve 9-u.c.-thick Au/YBCO nanowires at *T* = 4.2 K. IV curves of the nanowires oriented along a-axis, b-axis, and nodal directions are colored in green, blue, and red, respectively.

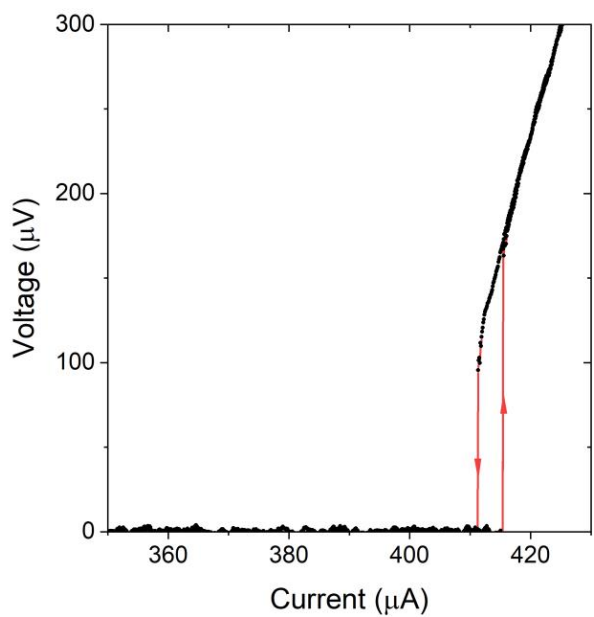

**Figure S2|** *IV* curve of 76-nm-wide and 9-u.c.-thick Au/YBCO nanowire at *T* = 4.2 K.

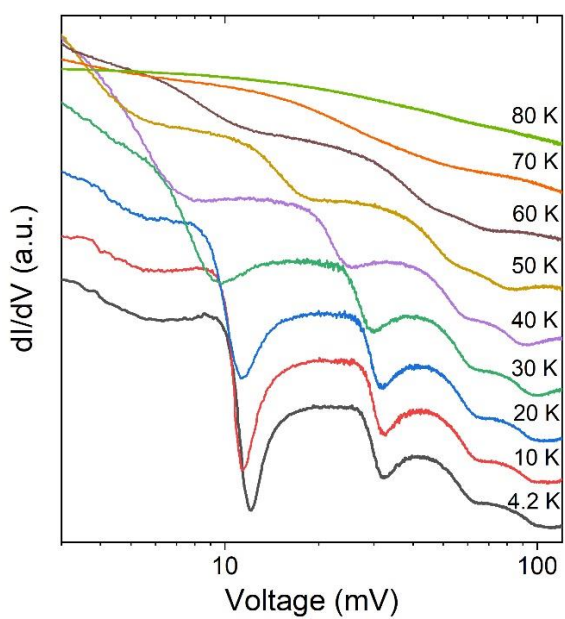

**Figure S3|** Differential conductance of the ultra-narrow 6-u.c.-thick YBCO nanoconstriction capped by the amorphous YBCO layer at various temperatures. Curves are shifted along the Y-axis for convenience of their representation.

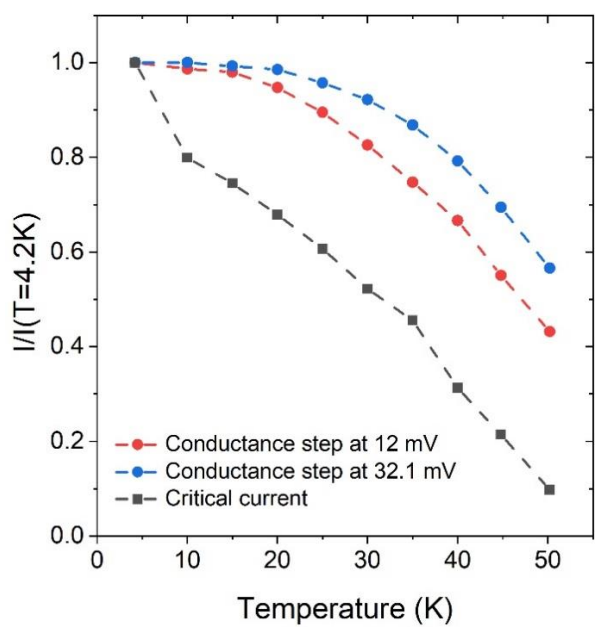

**Figure S4|** Temperature dependences of critical current and current position of the conductance steps at 12 mV and 32.2 mV of the ultra-narrow 6-u.c.-thick YBCO nanoconstriction capped by the amorphous YBCO.

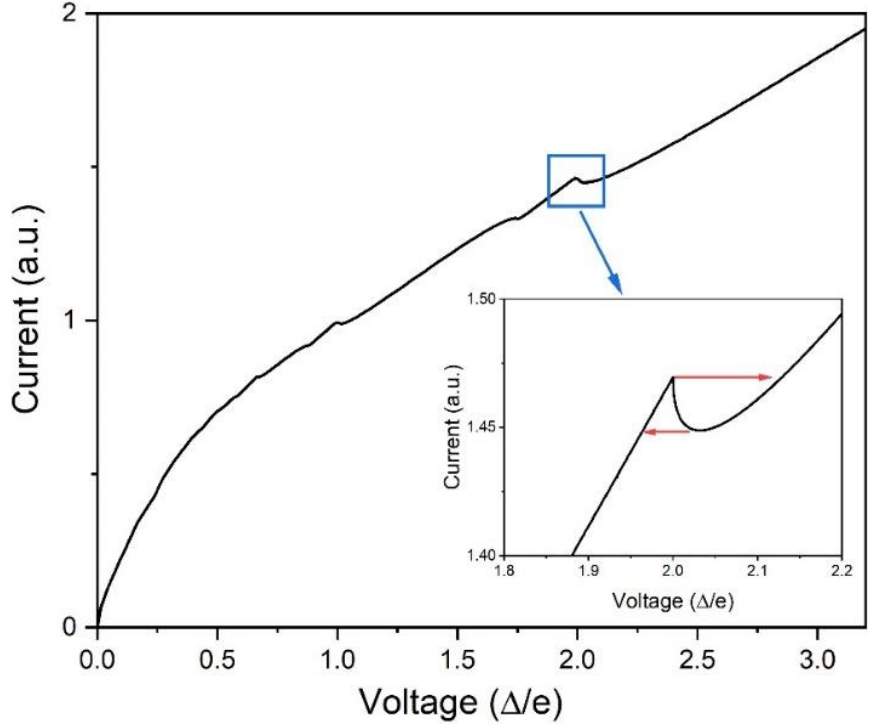

**Figure S5|** Numerically simulated current-voltage characteristic of the voltage-biased SNS junction. Inset. Zoomed current-voltage characteristic. Red arrows show the voltage switching in the current-bias regime.

**Structural properties of ultra-thin YBCO films**

To investigate the in-plane and out-of-plane crystallographic properties of the ultra-thin $YBa_2Cu_3O_{7-x}$ (YBCO) films, we deposited two 9.9-nm-thick YBCO films capped with a 5-nm-thick amorphous YBCO layer on the (100) $SrTiO_3$ substrate and perform X-ray diffraction (XRD) analysis with a high-resolution Rigaku Smartlab diffractometer. Both samples showed similar results. Figure S6a shows a high-resolution 2θ-θ scan of one of the samples. Only (00X) YBCO reflections are observed in the 2θ-θ scan, indicating single-crystalline growth along the c-axis as the growth direction. Additional reflections in the 2θ-θ scan correspond to $SrTiO_3$ substrate and CuO precipitates. An analysis of the scanning-electron microscopy micrographs shows that the precipitates have an average area of 0.45 $\mu m^2$ and occupy 0.36% of the film surface. From the diffraction angles of the (00X) peaks, we calculate the length of the YBCO c-axis parameter as 11.62 Å as shown in Figure S6b. High-resolution reciprocal space maps (RSMs) around the (0,1,8) and (1,0,8) reflections are shown in Figure S7. The peak around (0,1,8) reflection is very narrow and differs from the peak around (1,0,8) reflection. Optimally-doped YBCO has an orthorhombic unit cell with the table values of the lattice constants in a-, b-, and c- direction equal to 3.82, 3.89, and 11.67 Å, respectively. The b-axis is directed along the CuO chains. $SrTiO_3$ substrate has a cubic lattice cell at room temperature with a lattice constant of 3.905 Å. The misfit parameter for the a-axis direction of the YBCO is 2.2% while in the b-axis direction - only 0.39%. The smaller misfit parameter should correspond to the narrower RSM peak. Therefore, we have adopted the (1,0,8) and (0,1,8) asymmetric peaks to evaluate the in-plane lattices a- and b- of the YBCO film, respectively, that are assumed to be ordered along [100] and [010] crystal orientations on the SrTiO$_3$ substrate. The cross-sections of RSMs at the positions indicated by the green lines in Figure S7 are presented in Figure S8. We determine the a- and b-axis lattice parameters of the 9.9-nm-thick YBCO film from the peak positions of the RSM cross-sections as 3.88 and 3.90 Å, respectively, which means that the a- and b-axis lattice constants are nearly completely stressed to the SrTiO$_3$ lattice constant. The RSM peaks corresponding to the a- and b-axis lattice parameters of YBCO overlap which makes impossible an accurate calculation of the untwining degree. However, on the assumption that the small peak at 3.84 Å on the cross-section around (0,1,8) is due to the twin grains, we estimate the ratio of the twin grains as approximately 5% and the untwining degree as 95% in the film.

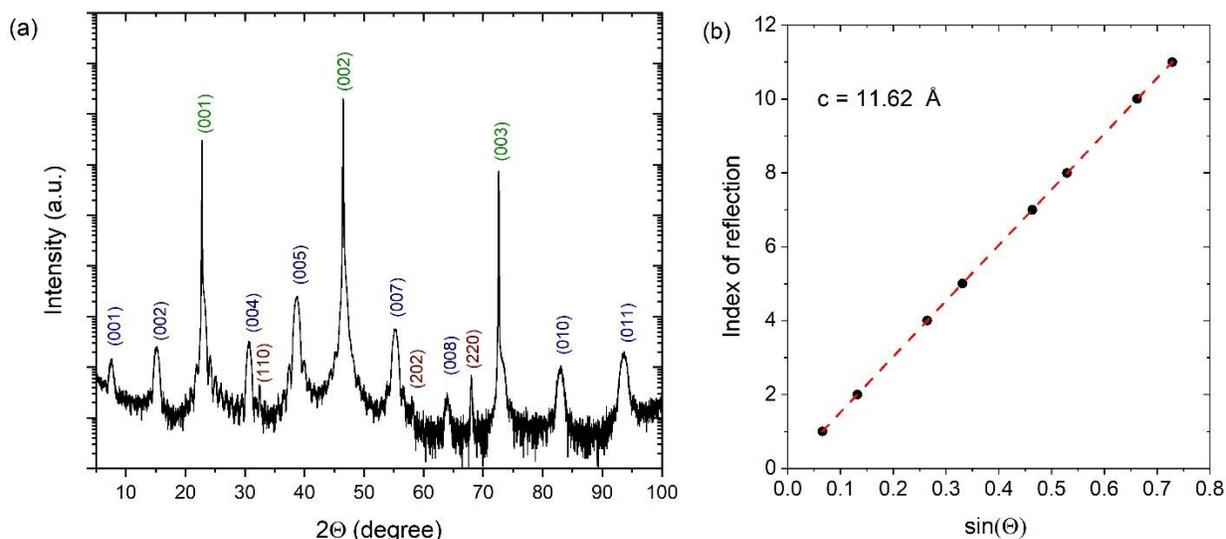

**Figure S6|** a. High-resolution XRD 2Θ- Θ scan of a 9.5-nm-thick YBCO film capped by a 5-nm-thick amorphous YBCO layer. Peaks corresponding to YBCO film, SrTiO$_3$ substrate, and CuO precipitates are colored in blue, green, and brown, respectively. b. YBCO film c-axis parameter obtained by the linear fitting of the (00X) reflections positions.

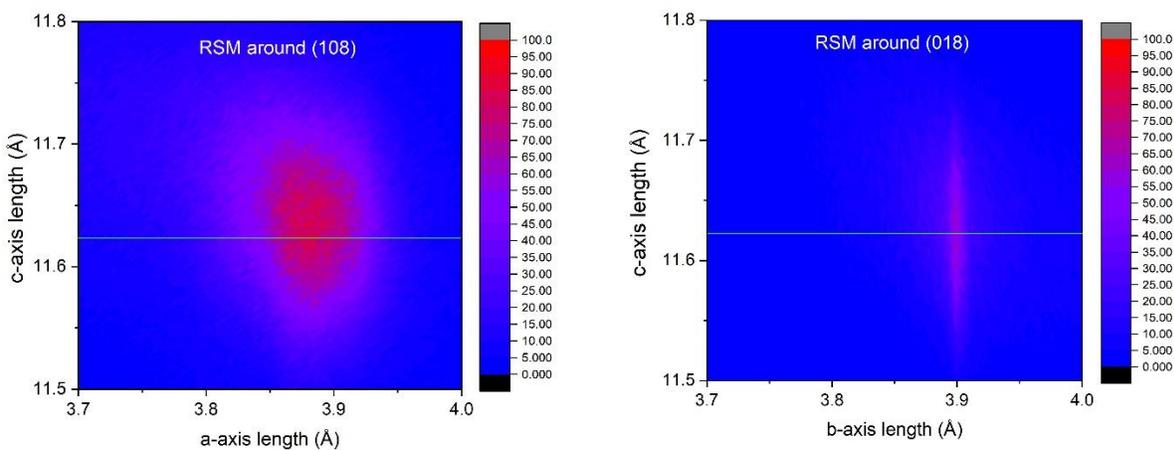

**Figure S7|** High-resolution reciprocal space maps (RSM) of a 9.5-nm-thick YBCO film capped by a 5-nm-thick amorphous YBCO layer around the (0,1,8) and (1,0,8) reflections. Position of the RSM cross section is shown by green line.

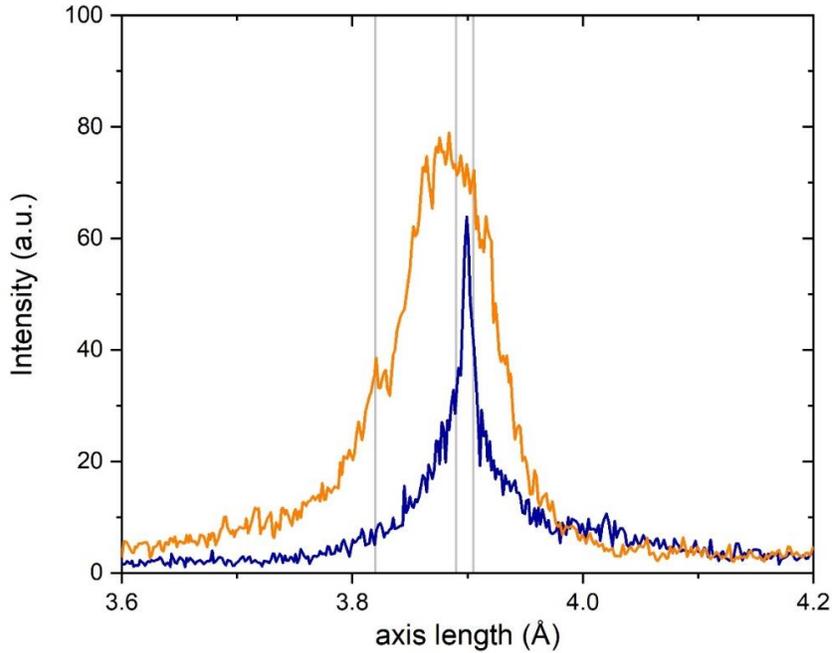

**Figure S8|** Cross sections of RSM around (108) reflection (orange line) and (018) reflection (blue line). Grey lines marks the table values of the a-axis and b-axis lattice constants of optimally-doped YBCO and the lattice constant of SrTiO$_3$ which are 3.82 Å, 3.89 Å, and 3.905 Å, respectively.

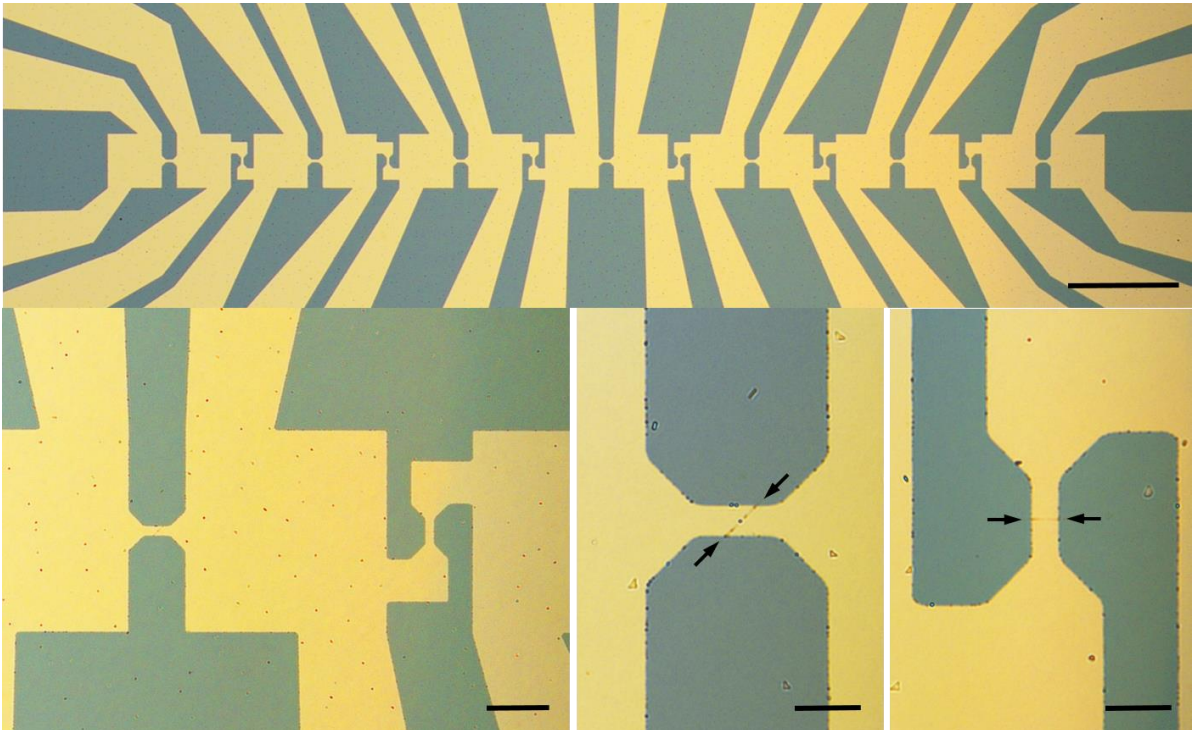

**Figure S9| Optical micrographs.** a. Sample layout. Scale bar is 200 µm. b. Microbridges aligned along different substrate edges. Scale bar is 30 µm. c. Zoomed image of the microbridge with "horizontal" orientation. Scale bar is 10 µm. d. Zoomed image of the microbridge with "vertical" orientation. Scale bar is 10 µm. The nanowire orientation is shown by black arrows.

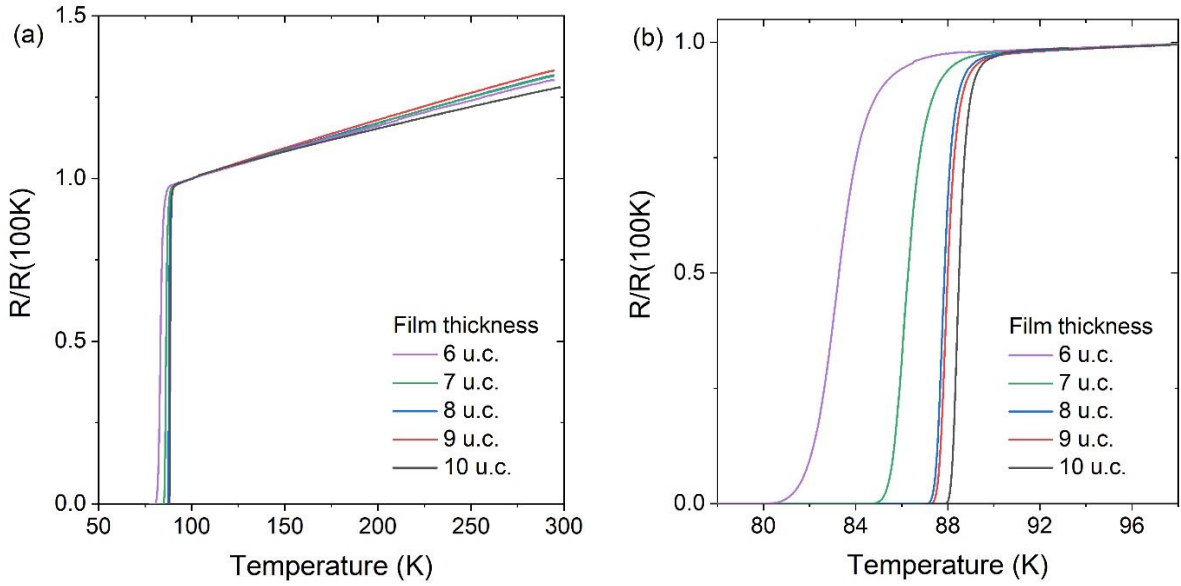

**Figure S10|** a. Temperature dependence of the normalized resistance of the 5-μm-wide Au/YBCO microbridges of various thicknesses. b. Zoomed temperature region of the superconducting transition of the Au/YBCO microbridges.

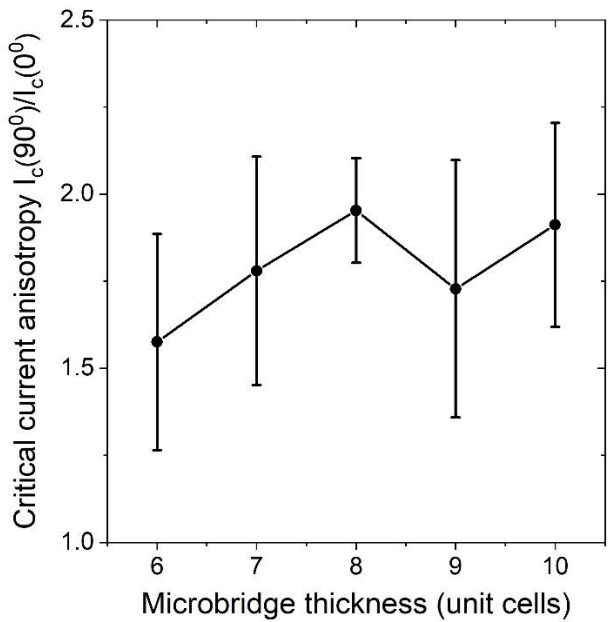

**Figure S11|** Anisotropy of the average critical current density <$J_{c90}/J_{c0}$> of the 5-μm-wide Au/YBCO microbridges of various thicknesses measured at temperature of 77.4 K.